\documentclass[journal]{IEEEtran}

\usepackage{subfigure}
\usepackage{color}
\usepackage{graphicx}
\usepackage{dcolumn}
\usepackage{bm}
\usepackage{physics}
\usepackage{blindtext}
\usepackage{float}
\usepackage{graphicx,wrapfig}
\newcommand*\diff{\mathop{}\!\mathrm{d}}
\newcommand{\vect}[1]{\mathbf{#1}}

\newcommand*\Laplace{\mathop{}\!\mathbin\bigtriangleup}
\newcommand*\Prob{P\left(r, t | r_0 \right)}

\ifCLASSINFOpdf
\else
\fi
\begin{document}
%
\title{Impulse Response of the Channel with a  Spherical Absorbing Receiver and a Spherical Reflecting Boundary}
%
%
%

\author{ Fatih Din\c{c},~\IEEEmembership{Student Member,~IEEE},
         Bayram Cevdet Akdeniz,~\IEEEmembership{Student Member,~IEEE}, Ali Emre Pusane,~\IEEEmembership{Member,~IEEE}  and Tuna Tugcu,~\IEEEmembership{Member,~IEEE} 

}

\maketitle

\begin{abstract}
In this letter, we derive the impulse response of the channel with a spherical absorbing receiver, a  spherical reflecting boundary, and a point transmitter in molecular communication domain. By exploring the channel characteristics and drawing comparisons with the unbounded case, we show the consequences of having the boundary on channel properties such as peak time, peak amplitude, and total fraction of molecules to hit the receiver. Finally, we calculate the bit error rate for both bounded and unbounded channels and emphasize the significance of incorporating the boundary on understanding the realistic behavior of a channel.
\end{abstract}


%
\IEEEpeerreviewmaketitle

\section{Introduction}

Nowadays, improvements in nanotechnology lead to reveal a new research area which is the communication of the nanomachines. Although many solutions are proposed in the literature, communication among nanomachines is still not extensively solved. Nonetheless, molecular communication via diffusion (MCvD) is one of the most promising approach due to its biocompability \cite{farsad2016comprehensiveSO}.

In order to investigate a communication system, modelling the received signal is an essential	phenomenon. In \cite{yilmaz2014arrival}, approximate distribution models for the received signal are presented in MCvD channels. Although the exact received signal model for a given transmission slot has binomial distribution, it is shown that it can also be approximated with Poisson and Gaussian distributions. On the other hand, in order to determine the parameters of the binomial (or Gaussian/Poisson) distribution, the channel impulse response is needed to determine the density of the received signal of a given time. In the literature, impulse response of many different channel models are derived by solving Fick's second law that defines the kinematics of  the diffusion using the necessary conditions that defines the corresponding channel.

These channel models can be categorized into various groups. Firstly, the receiver can  be either chosen as absorber that absorbs the incoming molecules and removes them from the environment  \cite{nakano2012channel} or as an observer that observes the incoming molecules without absorbing \cite{noel2014improving}. In general, the derivation of the channel model with an observing receiver is less challenging than the channel with an absorbing receiver, since an absorbing receiver violates the homogeneity of the environment, which this leads to a harder differential equation system. For instance, while the channel impulse response with a point transmitter and observing receiver is available for a 3-D environment without flow \cite{deng2015modeling} and with flow \cite{noel2014optimal}, the channel impulse response for an absorbing receiver is only derived for the without flow case as presented in \cite{yilmaz2014three}. Similarly the channel impulse response for the spherical receiver case has been analytically derived for both absorbing and observing receivers in \cite{noel2016channel}, while for the reflecting transmitter absorbing receiver case it can be solved by machine learning techniques \cite{genc2018reception}. 
Up to now, all channels in the literature are assumed to be unbounded. The only exception is the channel impulse response for tabular environments, which is considered in \cite{wicke2017modeling}. On the other hand, in practical situations, environments are generally bounded. 

Although one of the most frequently used models in the literature is the channel with a point transmitter and a spherical absorbing receiver, the bounded version of this channel is still an untouched area. In this paper, we derive the impulse response of the channel with a point transmitter and a spherical  absorbing receiver, whose center is concentric with a spherical reflective boundary. Once this model is derived and its verification is done with Monte Carlo simulations, we investigate the received signal properties and compare with the unbounded case by discussing their differences and similarities. Finally, the performances of the bounded and unbounded environments are compared.

\section{System Model}

\begin{figure}
    \centering
    \includegraphics[width=6cm]{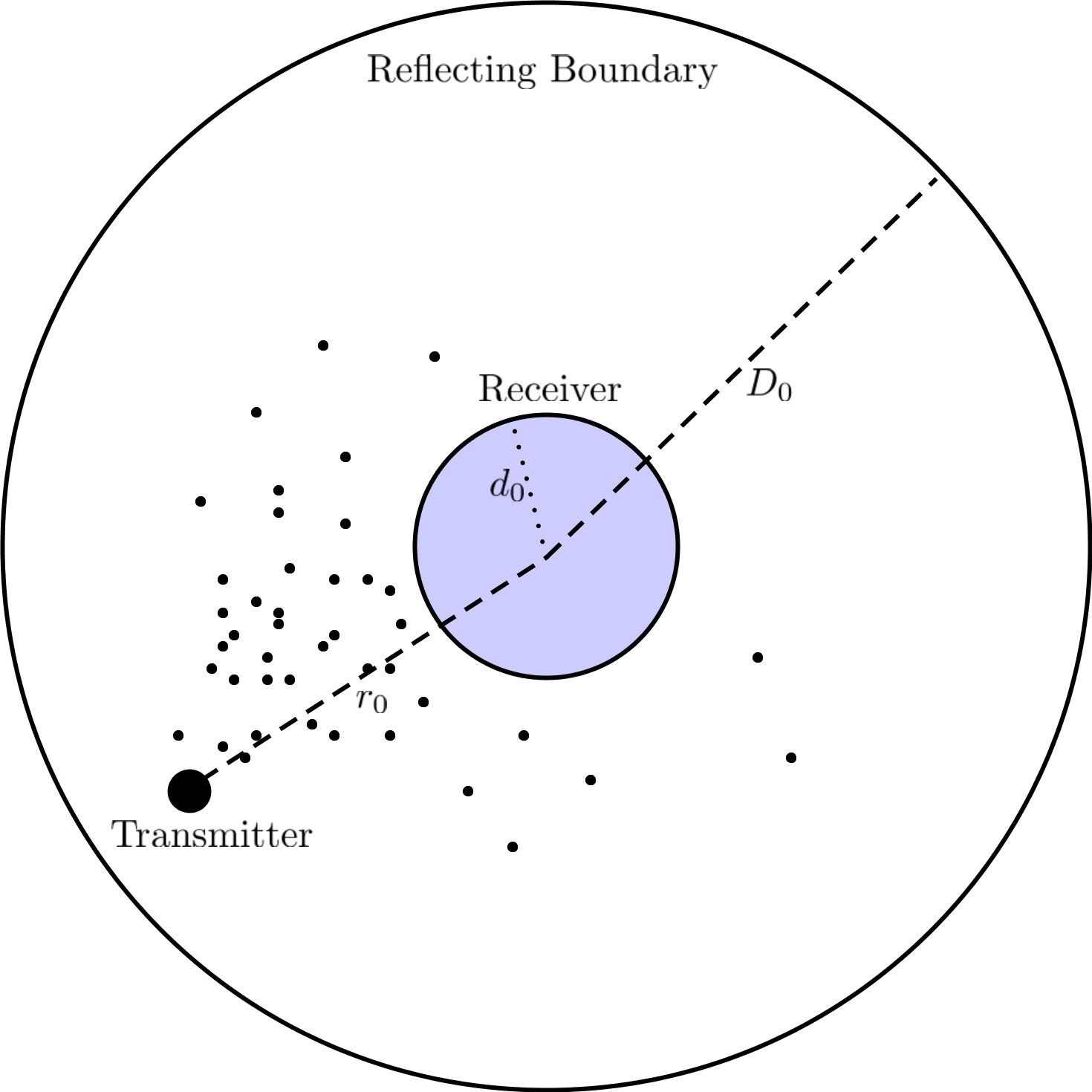}
    \caption{The system model for a channel consisting of a spherical receiver and a point transmitter with reflecting spherical boundaries. }
    \label{fig:main}
\end{figure}

The main purpose of this paper is to derive the impulse response for a spherical channel with a point transmitter as shown in Fig. \ref{fig:main}. In this channel, the transmission of a message carried by the messenger molecules is achieved through diffusion. For simplicity, we approximate the transmitter as a point source, where the initial distribution of a released molecule can be represented by a Dirac delta function ($\delta(r-r_0)/4\pi r^2$). With the assumption that there is no flow inside the medium, the diffusion of the molecules can be modelled as a Brownian motion process as
\begin{subequations} 
\begin{align}
  &\triangle x=\mathcal{N}\left( 0,2D\triangle t \right),\\
  &\triangle y=\mathcal{N}\left( 0,2D\triangle t \right),\\
  &\triangle z=\mathcal{N}\left( 0,2D\triangle t \right),
\end{align}
\end{subequations}\label{brown}
\noindent \hspace{-0.95em}
where $D$ is the diffusion coefficient and $\triangle t$ is the time step. The spherical receiver centered around the origin absorbs the molecules that reach its surface. If a molecule is incident upon the spherical boundary with radius $D_0$, it is reflected to the channel. 

\begin{figure*}
    \centering
    \includegraphics[width=9cm]{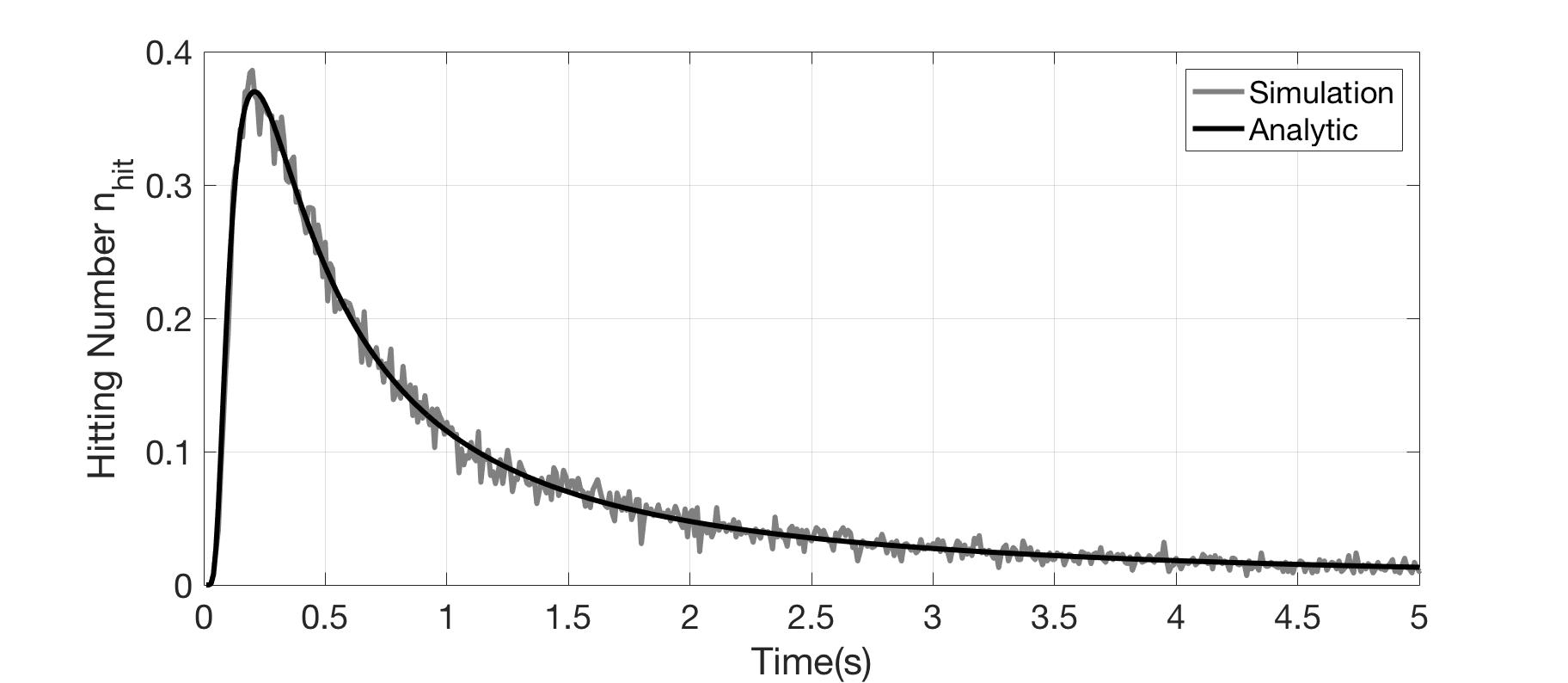}
    \includegraphics[width=9cm]{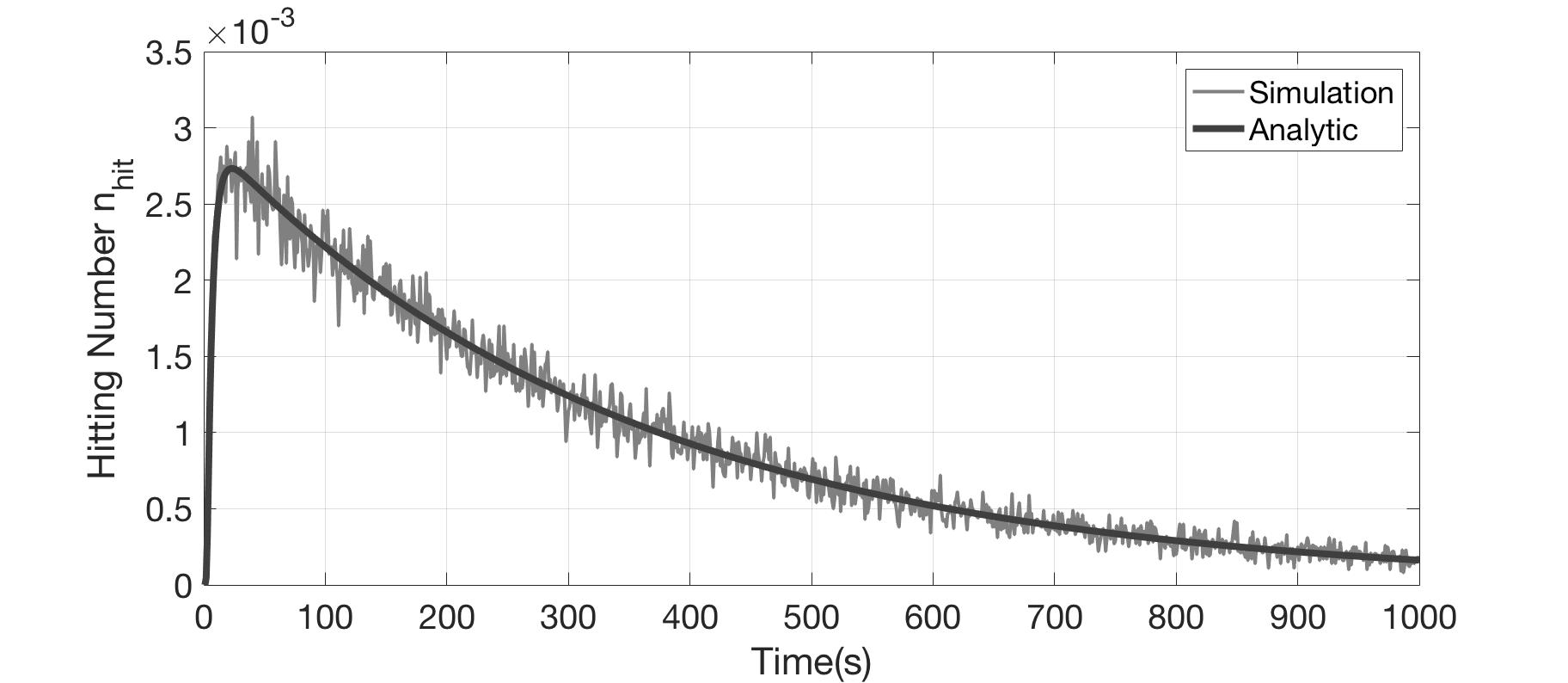}
    \caption{Simulation and analytic expression of the hitting number  $n_{hit}$ for a receiver with a radius $d_0=10um$, release point $r_0 = 20 um$ (left) and $r_0=80 um$ (right), boundary radius $D_0=100um$ and diffusion coefficient $D=80 um^2/s$ (right).} \label{fig:simsim}
\end{figure*}

\section{Channel Impulse Response}

The Fick's Law describes the diffusion of a molecule inside a region as
\begin{equation} \label{eq:fick}
D \grad^2 \Prob = \frac{\partial \Prob}{\partial t},
\end{equation}
where D is the diffusion coefficient, $\grad^2=\Laplace$ is the Laplacian and $\Prob$ is the probability density of the molecule. 

The boundaries reflect the incoming molecules, leading to the fact that the current of the molecules normal to the boundary is zero. Furthermore, the transmitter is assumed to be situated at a distance $r=r_0$ from the origin and in a completely angular symmetric manner in terms of the hitting probability. Finally, the probability distribution $\Prob$ should be zero when the molecules hit the receiver (assuming a perfect receiver due to simplicity), which result in the boundary conditions
\begin{subequations}  \label{eq:boundary}
\begin{align}
\frac{\partial \Prob}{\partial r}\Big|_{r=D_0} &= 0, \label{eq:Neuman} \\ 
 \Prob\Big|_{r=d_0} &= 0\label{eq:dirichlet},\\
 P(\vect r,0)&=\frac{1}{ 4 \pi r^2} \delta(r-r_0) .\label{eq:dirac} 
\end{align}
\end{subequations}

We start by the separation of variables as
\begin{equation}
\Prob=\psi(r,\theta,\phi) T(t)
\end{equation}
and obtain
\begin{equation*}
D \frac{\Laplace \psi(r,\theta,\phi)}{\psi(r,\theta,\phi)} = \frac{T'(t)}{T(t)}= - \mu^2. 
\end{equation*}
from which we can easily deduce that
\begin{equation*}
T(t) = e^{-\mu^2 t}.
\end{equation*}
This leads to the eigenvalue problem for Laplacian operator
\begin{equation}
 \Laplace \psi(r,\theta,\phi) = - \frac{\mu^2}{D} \psi(r,\theta,\phi).
\end{equation}

At this point, let us define $\psi (r,\theta,\phi)=R(r)$ by exploiting the SO(3) symmetry of the system and open up the Laplacian in spherical coordinates to find
\begin{align*}
r^2R''(r) + 2r R' + \frac{\mu^2}{D} r^2  R =0. 
\end{align*}
The solution to this equation can be written as
\begin{equation*}
R(r) = j_0(\beta_n r/D_0) + c n_0 (\beta_n r/D_0),
\end{equation*}
where $\beta_n = \frac{\mu D_0}{\sqrt{D}}$. Let us now define the function
\begin{equation}
\kappa_0(\beta_n x) = j_0(\beta_n x) + c_n y_0(\beta_n x)
\end{equation}
such that $\kappa_0(\beta_n)'=0$ and $\kappa_0(\beta_n \alpha)=0$ ($\alpha=d_0/D_0$). Then, the function $\kappa_0(\beta_n r/D_0)$ satisfies the boundary conditions and is a solution to the Fick's Law. Then, the most general probability density function can be obtained as
\begin{align}
\Prob = \sum_n A_n \kappa_0(\beta_n r/D_0) e^{-\beta_n^2 \frac{D t}{D_0^2}}
\end{align}

Before finding the expansion coefficient $A_n$, we use straightforward algebra to obtain the following orthogonality condition
\begin{align*}
\int_\alpha^1 \diff x x^2 \kappa_0(\beta_{n} x) \kappa_{0} (\beta_{n'}x) = \frac{\pi}{2\beta_n} \Big( \frac{x^2}{2} (\eta_{1/2}(\beta_{n} x)^2  \\  - \eta_{3/2}(\beta_{n}x) \eta_{-1/2}(\beta_{n}x)  ) \Big)\Bigg|_\alpha^1  \delta_{n n'}  = I_n \delta_{n n'},
\end{align*}
where $\eta_m(\beta_n x) = J_m(\beta_n x) + c_n Y_m(\beta_n x)$  are the special functions defined using Bessel functions of the first and second kind. Thus, the probability distribution function (PDF) can be rewritten as
\begin{align}
\Prob= \sum_n \frac{\kappa_0(\beta_n r_0/D_0)}{4 \pi I_n D_0^3} \kappa_0(\beta_n r/D_0) e^{-\beta_n^2 \frac{D t}{D_0^2}}.
\end{align}
Having found the PDF, we can define the hitting number as
\begin{equation}
    n_{hit} = 4 \pi r^2 D \partial_r \Prob |_{r=d_0},
\end{equation}
which results in
\begin{align}
    n_{hit}&= \sum_n \frac{\beta_n \alpha^2 }{ I_n} \frac{D}{D_0^2} \kappa_0(\beta_n r_0/D_0)\kappa_0'(\beta_n \alpha) e^{-\beta_n^2 \frac{D t}{D_0^2}},
\end{align}
where $\kappa_0'(x)$ denotes the derivative of $\kappa_0(x)$ with respect to $x$. An illustration of the equivalence between our analytical result and the Monte-Carlo simulations can be found in  \\ Fig. \ref{fig:simsim}.

By integrating the hitting number, we can find the number of particles received after time $t$ as
\begin{equation}
    N_{tot}(t) = \int_0^\tau n_{hit}(\tau) \diff \tau,
\end{equation}
which is given as
\begin{equation} \label{eq:hittot}
    N_{tot} (t) = \sum_n \frac{\alpha^2 \kappa_0(\beta_n r_0/D_0) \kappa_0'(\beta_n \alpha) \left(1-\exp(-\beta_n^2 \frac{D t}{D_0^2}) \right)}{I_n \beta_n}.
\end{equation}
We note that, up to this point, we haven't made any approximations yet. Due to the exponential time dependence of this sum, there will be a leading term for large values of $t$. Thus, we can approximate the hitting probability for large $t$ as
\begin{equation*}
    N_{tot} (t) \simeq 1- \frac{\alpha^2}{I_1 \beta_1} \kappa_0(\beta_1 r_0/D_0) \kappa_0'(\beta_1 \alpha) \exp(-\beta_1^2 \frac{D t}{D_0^2}),
\end{equation*}
from which we find the $\epsilon$ fraction of the molecules not absorbed after time $t$ by
\begin{equation*}
    \epsilon = \frac{\alpha^2}{I_1 \beta_1} \kappa_0(\beta_1 r_0/D_0) \kappa_0'(\beta_1 \alpha) \exp(-\beta_1^2 \frac{D t}{D_0^2}).
\end{equation*}
Thus, for a given $\epsilon$, we can find the time $t^*$ that $1-\epsilon$ fraction of the molecules are absorbed as
\begin{equation}
    t^* = \frac{D_0^2}{\beta_1^2 D} \ln(\frac{\alpha^2 \kappa_0(\beta_1 r_0 /D_0) \kappa_0'(\beta_1\alpha)}{\epsilon I_1 \beta_1}).
\end{equation}
Here, we can realize that $\left|\ln(\frac{\alpha^2 \kappa_0(\beta_1 r_0 /D_0) \kappa_0'(\beta_1\alpha)}{I_1 \beta_1})\right|<1$, from which we can approximate the hitting time as:
\begin{equation} \label{eq:tmax}
    t^*_{max} \simeq -\frac{D_0^2}{\beta_1^2 D} \ln \epsilon.
\end{equation}

\begin{figure}
    \centering
    \includegraphics[width=8cm]{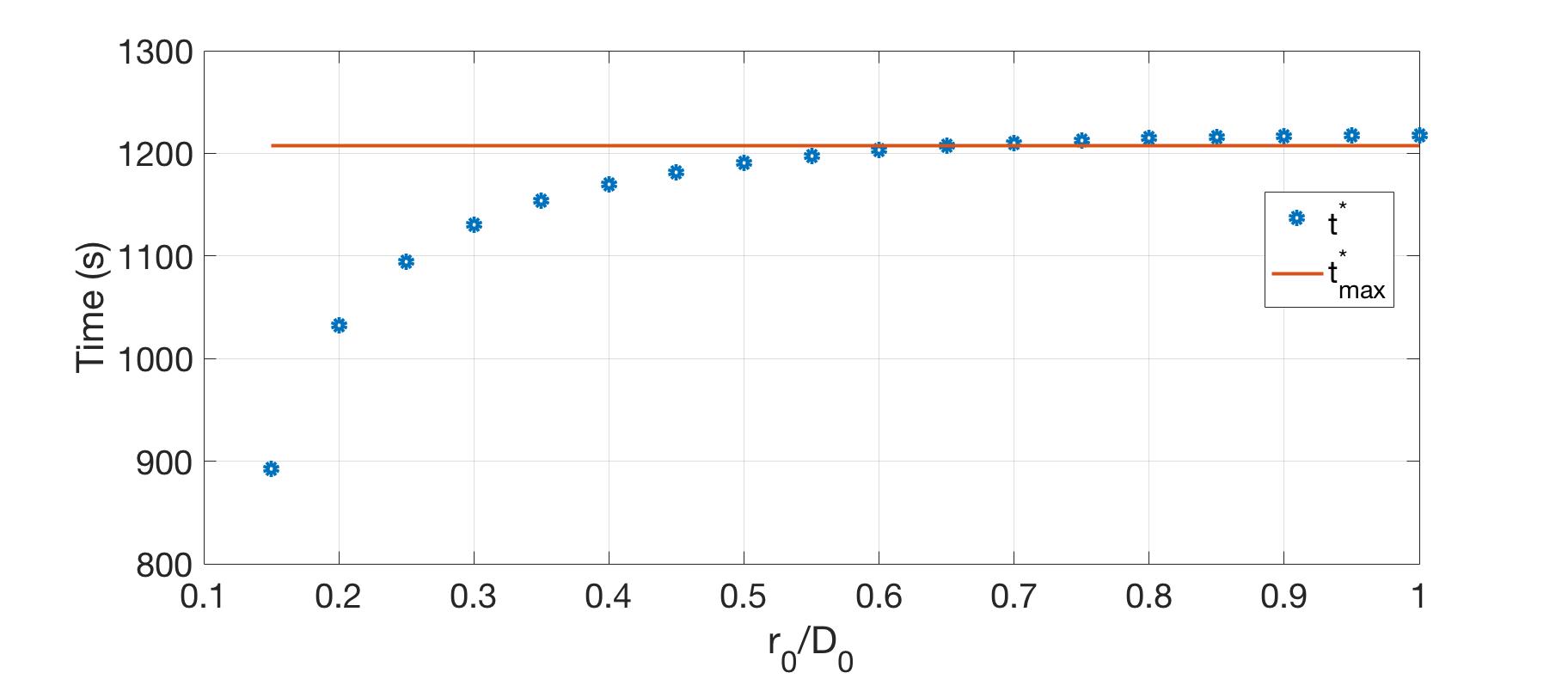}
    \caption{The dependence of hitting time $t^*$ on the initial distance $r_0$ of the transmitter with $D=80\mu m^2/s$, $D_0=100\mu m$, $d_0=10\mu m$ and $\epsilon=0.03$.} \label{fig:hittime}
\end{figure}

Moreover, we can plug time $t^*$ into  (\ref{eq:hittot}) to realize that the second leading term scales as $\sim \epsilon^{(\beta_2/\beta_1)^2}$, where $(\beta_2/\beta_1)^2\simeq 69.3$. This ensures that, for arbitrarily small $\epsilon$, $t^*$ is indeed the time after which $1-\epsilon$ fraction of the molecules is absorbed by the receiver. An illustration of $t^*$ and $t^*_{max}$ is given in \\ Fig. \ref{fig:hittime}.

\section{Molecular Communication Characteristics of the Channel}

In this section, we explore the effects of the reflecting boundary on the molecular communication characteristics of a spherical receiver and expand on the findings of \cite{yilmaz2014three} regarding the unbounded receiver. In this section, we shall discuss the peak time $\tau_{peak}$ and the peak amplitude $n_{peak}$, as well as their dependency on the boundary and the initial point $r_0$ of the transmitter.

\begin{figure}
    \centering
    \includegraphics[width=8cm]{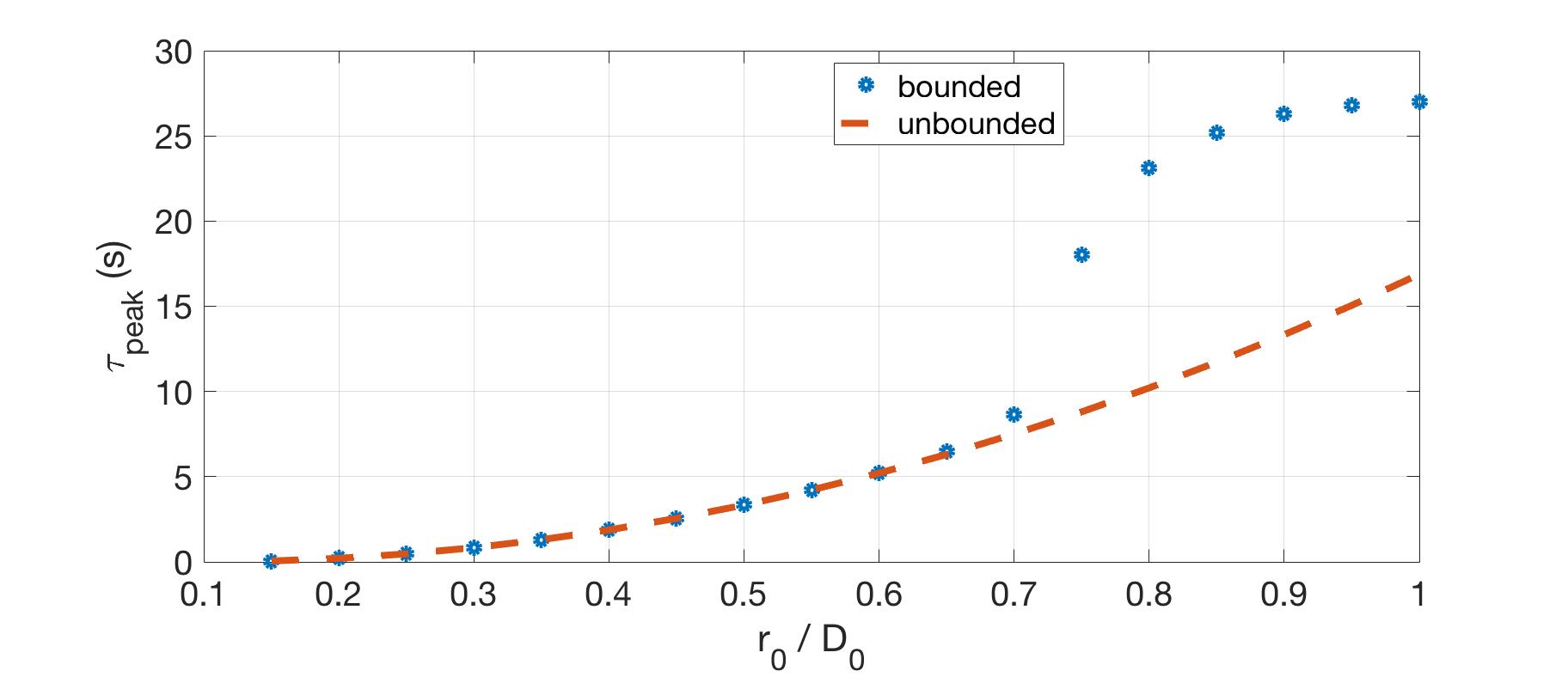}
    \caption{The peak time $\tau_{peak}$ for the bounded and unbounded channels vs. $r_0/D_0$, where $D_0=100\mu m$ and $d_0=10\mu m$. For the unbounded channel, $\tau_{peak} = d^2/6D$ \cite{yilmaz2014three}. Note the dominant effect of the boundary for $(r_0-d_0) \simeq 2/3 l_c$, where $l_c=D_0-d_0$ is defined as the channel length for this case, around $r_0 \simeq 66 \mu m$. Comparing with Fig. \ref{fig:hittime}, one can realize the dominating effect of the tail on $N_{tot}(t)$, as $t^*>> \tau_{peak}$.} \label{fig:time}
\end{figure}

\begin{figure}
    \centering
    \includegraphics[width=8cm]{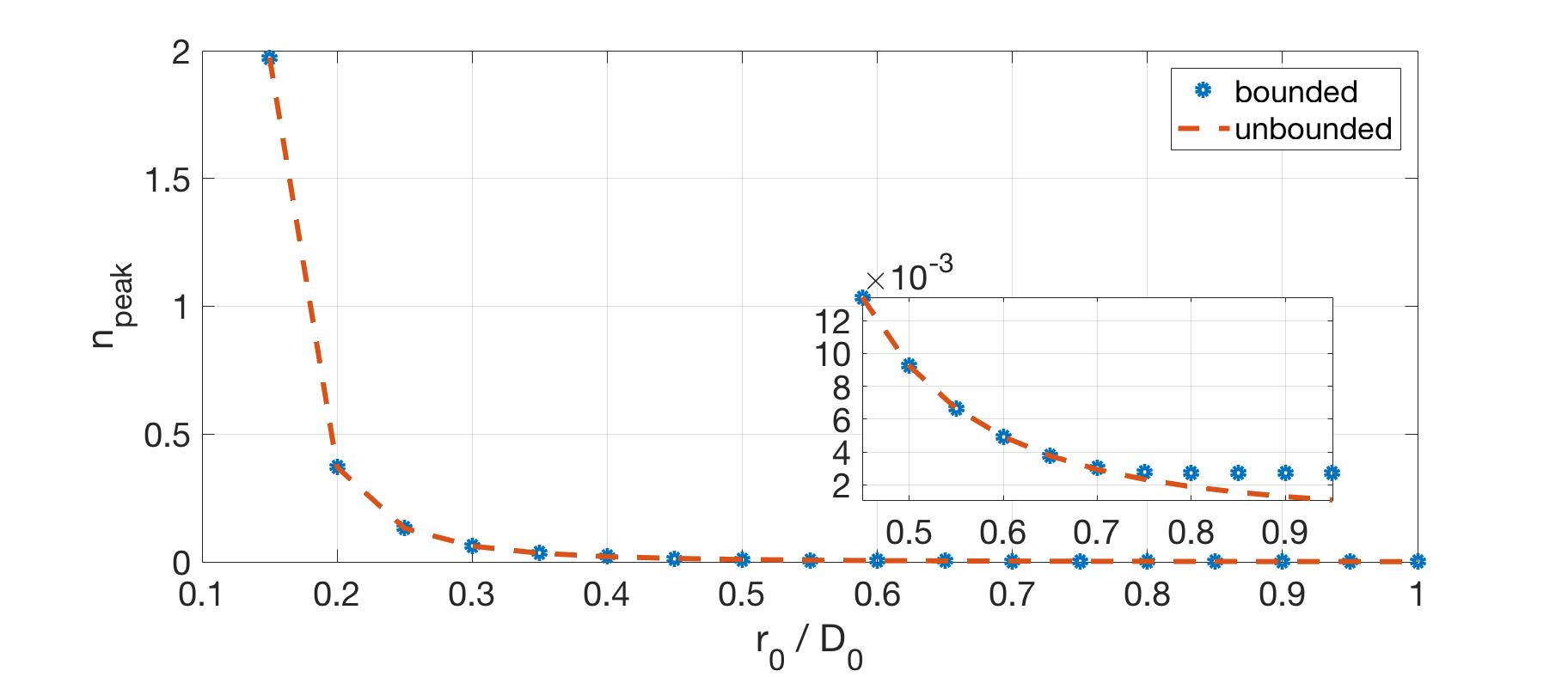}
    \caption{The peak amplitude $n_{peak}$ for the bounded and unbounded channels vs $r_0/D_0$, where $D=80\mu m^2/s$, $D_0=100\mu m$ and $d_0=10\mu m$. For the unbounded channel, $n_{peak} = (d_0 D e^{-3/2})/(r_0 (r_0-d_0)^2 \sqrt{\pi/54})$ \cite{yilmaz2014three}. Note the dominant effect of the boundary for $(r_0-d_0) \simeq 2/3 l_c$, where $l_c=D_0-d_0$ is defined as the channel length. (For this case, around $r_0 \simeq 66 \mu m$)} \label{fig:n}
\end{figure}
\begin{figure*}
\centering

\subfigure[ $D=80 { \mu { m }^{ 2 } }/{ s }$ ]{
\includegraphics[width=.48\textwidth]{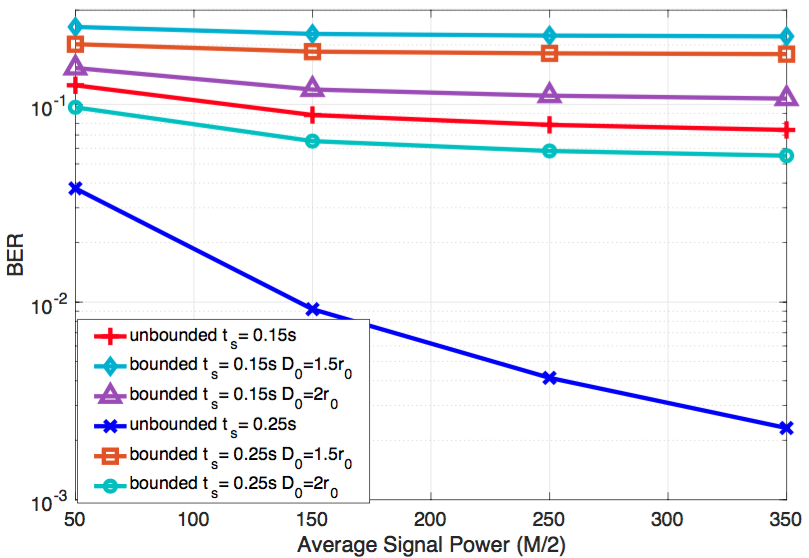}
\label{a1}
}
\subfigure[ $D=800 { \mu { m }^{ 2 } }/{ s }$ ]{
\includegraphics[width=.48\textwidth]{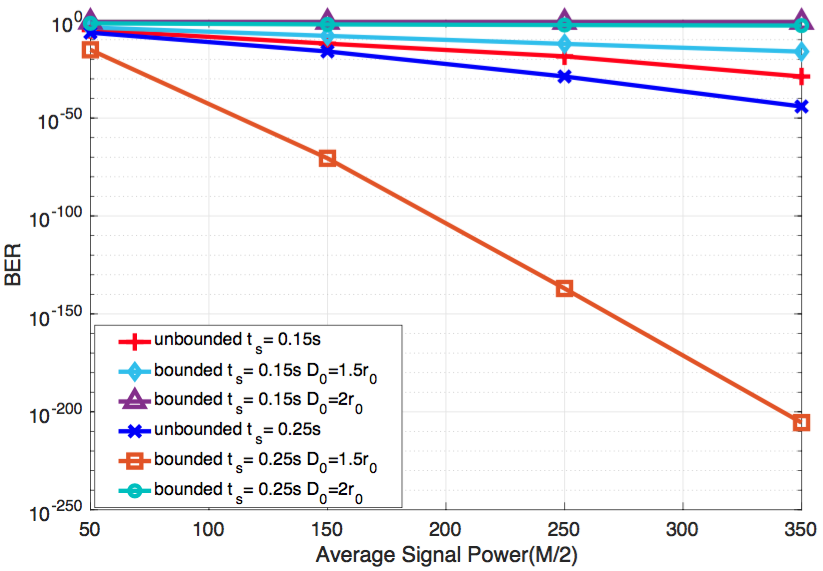}
\label{a2}
}
\caption{BER comparison of the bounded and unbounded channels with different parameters and diffusion coefficient $D=80 um^2/s$ (left) and $D=800 um^2/s$ (right). For the simulations, $r_0=10um$, $d_0=5um$ and $D_0$ is varied accordingly. In the presence of the boundary, the fraction of particles absorbed by the receiver after the time $t_s$ is given as $\epsilon \sim \exp(-\beta_1^2 D t_s/D_0^2 )$ from (\ref{eq:tmax}). For $D = 800 um^2/s$, the fraction of absorbed molecules is $0.85$ for $D_0=1.5r_0$ and $0.65$ for $D_0=2r_0$.}
\label{fig:ber}
\end{figure*}

First, let us define the channel length parameter as $l_c=D_0 - d_0$. This parameter defines the scale of the channel , which we select as $l_c=90 \mu m$ for our simulations. Then, we define the distance of the transmitter $d=r_0-d_0$. Comparing $\tau_{peak}$ in bounded and unbounded channels, we obtain the results presented in Fig. \ref{fig:time}. We note that, at about $d \simeq 2/3 l_c$, there is an apparent trend shift in the behavior between the bounded and the unbounded channels. This trend shift can also be seen from Fig. \ref{fig:n}.

In order to examine the effects of boundary clearly, it is beneficial to compare the bit error rate (BER) performances of the same environment with and without a boundary. In Fig. \ref{fig:ber}, BER curves of the bounded and unbounded channels are presented. As can be seen in Fig. \ref{a1}, the unbounded channel has better performance than the bounded case. This is an interesting but expected result. It is interesting, since all molecules are absorbed in the bounded case while in the unbounded case only $\frac{d_0}{r_0}$ of the molecules are absorbed. The reason of this difference is the reflective property of the environment that leads to direct the unabsorbed molecules to the receiver. On the other hand, since it can take  much time for a molecule to be  reflected by the boundary and subsequently be absorbed by the receiver, the molecules reflected from the boundary and then absorbed by the receiver most probably belong to the previous symbols; hence, they tend to lead to inter symbol interference. Furthermore, for the same reason, BER increases in the bounded cases as boundary radius $D_0$ decreases. On the other hand, as seen in the Fig. \ref{a2} that involves a 10 times higher diffusion coefficient $D$ than the one whose results are presented in Fig. \ref{a1}, it can be observed that for some cases, the bounded environment  may have far more successful performance than the unbounded case. This case occurs if $t_s$ gets closer to  $t^*_{max}$, which decreases as $D$ increases and $D_0$ decreases. Therefore, it is not surprising that  BER is directly proportional to $D_0$, unlike for the cases in \\ Fig. \ref{a1}.

\section{Conclusion}

In this letter, we have presented a bounded molecular communication channel consisting of a boundary, a point transmitter, and a spherical receiver. We model the boundary as a reflecting sphere and derive the impulse response for this channel. Our calculations show that the boundary forces all molecules to be absorbed by the receiver, in contrary to the unbounded case where only a $d_0/r_0$ fraction is absorbed \cite{yilmaz2014three}. Nonetheless, this happens in a relatively long time, approximately $50 \tau_{peak}$. Depending on the time $t_s$, MCvD can either benefit from (for large $t_s$) or get disturbed by the existence of the boundary. A reasonable value for $t_s$ can be determined if the channel dimensions are in nano-scales. Finally, through BER comparison, we show that the tail effect resulting from the previously released molecules makes the existence of the boundary undesirable for molecular communication purposes in micro or larger dimensions, except for molecules with extremely high diffusion coefficient $D$ values.

 \bibliographystyle{IEEEtran}
\bibliography{ref}

\begin{thebibliography}{10}
\providecommand{\url}[1]{#1}
\csname url@samestyle\endcsname
\providecommand{\newblock}{\relax}
\providecommand{\bibinfo}[2]{#2}
\providecommand{\BIBentrySTDinterwordspacing}{\spaceskip=0pt\relax}
\providecommand{\BIBentryALTinterwordstretchfactor}{4}
\providecommand{\BIBentryALTinterwordspacing}{\spaceskip=\fontdimen2\font plus
\BIBentryALTinterwordstretchfactor\fontdimen3\font minus
  \fontdimen4\font\relax}
\providecommand{\BIBforeignlanguage}[2]{{%
\expandafter\ifx\csname l@#1\endcsname\relax
\typeout{** WARNING: IEEEtran.bst: No hyphenation pattern has been}%
\typeout{** loaded for the language `#1'. Using the pattern for}%
\typeout{** the default language instead.}%
\else
\language=\csname l@#1\endcsname
\fi
#2}}
\providecommand{\BIBdecl}{\relax}
\BIBdecl

\bibitem{farsad2016comprehensiveSO}
N.~Farsad, H.~B. Yilmaz, A.~Eckford, C.-B. Chae, and W.~Guo, ``A comprehensive
  survey of recent advancements in molecular communication,'' \emph{{IEEE}
  Commun. Surveys Tuts.}, vol.~18, no.~3, pp. 1887--1919, 2016.

\bibitem{yilmaz2014arrival}
H.~B. Yilmaz and C.-B. Chae, ``Arrival modelling for molecular communication
  via diffusion,'' \emph{Electronics Letters}, vol.~50, no.~23, pp. 1667--1669,
  2014.

\bibitem{nakano2012channel}
T.~Nakano, Y.~Okaie, and J.-Q. Liu, ``Channel model and capacity analysis of
  molecular communication with {Brownian} motion,'' \emph{IEEE Communications
  Letters}, vol.~16, no.~6, pp. 797--800, 2012.

\bibitem{noel2014improving}
A.~Noel, K.~C. Cheung, and R.~Schober, ``Improving receiver performance of
  diffusive molecular communication with enzymes,'' \emph{IEEE Transactions on
  NanoBioscience}, vol.~13, no.~1, pp. 31--43, 2014.

\bibitem{deng2015modeling}
Y.~Deng, A.~Noel, M.~Elkashlan, A.~Nallanathan, and K.~C. Cheung, ``Modeling
  and simulation of molecular communication systems with a reversible
  adsorption receiver,'' \emph{IEEE Transactions on Molecular, Biological and
  Multi-Scale Communications}, vol.~1, no.~4, pp. 347--362, 2015.

\bibitem{noel2014optimal}
A.~Noel, K.~C. Cheung, and R.~Schober, ``Optimal receiver design for diffusive
  molecular communication with flow and additive noise,'' \emph{IEEE
  transactions on nanobioscience}, vol.~13, no.~3, pp. 350--362, 2014.

\bibitem{yilmaz2014three}
H.~B. Yilmaz, A.~C. Heren, T.~Tugcu, and C.-B. Chae, ``Three-dimensional
  channel characteristics for molecular communications with an absorbing
  receiver,'' \emph{IEEE Communications Letters}, vol.~18, no.~6, pp. 929--932,
  2014.

\bibitem{noel2016channel}
A.~Noel, D.~Makrakis, and A.~Hafid, ``Channel impulse responses in diffusive
  molecular communication with spherical transmitters,'' \emph{in Proc. CSIT
  Biennial Symposium on Communications Jun. 2016}.

\bibitem{genc2018reception}
G.~Genc, Y.~E. Kara, T.~Tugcu, and A.~E. Pusane, ``Reception modeling of
  sphere-to-sphere molecular communication via diffusion,'' \emph{Nano
  Communication Networks}, 2018.

\bibitem{wicke2017modeling}
W.~Wicke, T.~Schwering, A.~Ahmadzadeh, V.~Jamali, A.~Noel, and R.~Schober,
  ``Modeling duct flow for molecular communication,'' \emph{arXiv preprint
  arXiv:1711.01479}, 2017.

\end{thebibliography}

%




\end{document}